# Role of charge doping and lattice distortions in codoped Mg$_{1-x}$(AlLi)$_x$B$_2$ compounds.

M.Monni C.Ferdeghini and M.Putti
CNR-INFM-LAMIA and Dipartimento di Fisica, Università di Genova, Via Dodecaneso 33, 16146 Genova, Italy
E-mail: monni@fisica.unige.it

P.Manfrinetti A.Palenzona
CNR-INFM-LAMIA and Dipartimento di Chimica e Chimica Industriale, Università di Genova, Via Dodecaneso 31, 16146 Genova, Italy

M.Affronte
CNR-INFM-S3 and Dipartimento di Fisica, Università di Modena e Reggio Emilia
Via G.Campi 213/A, I-41100 Modena, Italy

P.Postorino, M.Lavagnini, A.Sacchetti, D.Di Castro
CNR-INFM-COHERENTIA and Dipartimento di Fisica, Università di Roma "La Sapienza", Piazzale A. Moro 2, 00185 Roma, Italy

F.Sacchetti, C.Petrillo, A.Orecchini
CNR-INFM CRS-SOFT and Dipartimento di Fisica, Università di Perugia, I-06123, Perugia, Italy

We prepared a series of Mg$_{1-x}$(AlLi)$_x$B$_2$ samples with 0≤x≤0.45 in order to compensate with Li the electron doping induced by Al. Structural characterization by means of neutron and X-ray diffraction confirms that Li enters the MgB$_2$ structure even though in an amount less than nominal one. We performed susceptibility, resistivity and specific heat measurements. Vibrational properties were also investigated by means of Raman spectroscopy. We compare these results with those obtained on a homologous series of Mg$_{1-x}$Al$_x$B$_2$ samples. The systematic success of scaling the relevant properties with the Al content rather than with the electron doping suggests that lattice deformation plays an important role in tuning the superconducting properties.

# 1. Introduction

The discovery of superconductivity in MgB$_2$ with a critical temperature T$_c$ = 39K renewed the interest for novel effects in two-gap superconductors. *Ab initio* calculations [1,2,3] showed that MgB$_2$ is characterized by two weakly coupled gaps $\Delta_\sigma(0) \approx$ 7 meV and $\Delta_\pi(0) \approx$ 2 meV residing on disconnected sheets of the Fermi surface formed by in-plane p$_{xy}$ boron orbitals ($\sigma$ band) and out-of-plane p$_z$ boron orbitals ($\pi$ band). The two-gap Eliashberg theory has explained most of the anomalies in superconducting and normal properties of pure MgB$_2$. However, the physics of two-gap MgB$_2$ alloys is still poorly understood.

Chemical substitution effects have been one of the hot topics since the beginning. Despite its simple structure and apparently simple chemistry, MgB$_2$ has so far proved very difficult to modify systematically through chemical substitutions. Various substitutions have been reported, but only a few were successful [4]. These are the cases of Al on Mg site [4-10] and of C on B site [11-16]. The primary effect of Al-/C-substitution is a drop of T$_c$ which can be ascribed to several different possible sources such as a decrease in the density of states and the weakening of electron phonon coupling. Moreover the activation of inter-band scattering with nonmagnetic impurities mixes $\sigma$ and $\pi$ Cooper pairs, and, averaging the order parameters reduces T$_c$ down to the isotropic value [17].

The role of charge doping and lattice distortions in alloyed MgB$_2$ is largely debated. Recently J.Kortus and co-workers [21] considered data of Al and C doped samples from several groups and explained the T$_c$ reduction mainly as an effect of band filling. Moreover, in order to explain the behaviour of the gaps with doping, they proposed that inter-band scattering effects are more important in C doped than in Al doped samples. This suggestion opened a lively debate [22,23]. In fact, it contradicts previous calculations [20] that predict relevant inter-band scattering be enhanced by c-axis displacements produced by substitution of Mg with atoms of different size, which is the case of Al, and not by substitution of B with C. Finally, a strong T$_c$ suppression has been observed in irradiated samples [24, 25, 26] where charge doping can be excluded. The merging of the gaps recently observed in irradiated samples [27] proves the importance of inter-band scattering even if a role of intra-band scattering should be envisaged. To clarify the role of disorder-induced inter-band scattering S.C.Erwin and I.I.Mazin [20] suggested a co-doping with Na and Al in the Mg sites. Such a compound, isoelectronic with $MgB_2$, should present significant interband scattering due to lattice distortions induced by substitutions of Al and Na in Mg sites and no charge doping under the assumption that the valence of dopant can be simply summed. Following this suggestion, recently renewed also by J.Kortus [23], we investigated the compounds $Mg_{1-x}(AlLi)_xB_2$

We chose Li instead of Na, since, although they have the same valence, Li shows a more favourable packaging effect owing to its atomic radius which is comparable with that of Mg.

Several compounds of this family were first synthesized by G.J. Xu and co-workers [28]. In this paper we present the preparation of $Mg_{1-x}(AlLi)_xB_2$ polycrystalline samples with x=0-0.45. Due to the difficult detection of Li, a careful crystallographic characterization by x-ray (XRD) and neutron diffraction was performed. Normal and superconducting properties were studied by resistivity, susceptibility and specific heat measurements and the phonon spectrum was analyzed by Raman spectroscopy. The results are compared with those obtained on the homologous series of $Mg_{1-x}Al_xB_2$ samples, with the aim of clarifying the role of disorder, and electron doping on the physical properties of $MgB_2$. Our findings suggest that disorder more than the electron doping, plays an important role for the suppression of superconductivity in $MgB_2$.

## 2. Sample preparation

Two series of $Mg_{1-x}(AlLi)_xB_2$ samples were prepared by direct synthesis of pure elements, the first with natural Boron (0 ≤ x ≤ 0.45), and the second with isotopically enriched $^{11}B$ (x=0, 0.1, 0.2, 0.3) in order to carry out neutron measurements. The AlLi alloy was firstly prepared and mixed with Mg with concentration ranging from 0 to 0.45. The Mg-(AlLi) mixture was then pressed in a pellet and put together with crystalline B in Ta crucibles, welded in argon and closed in quartz tubes under vacuum. The quartz tube was then placed in a vertical furnace at 850-900°C. The Mg-(AlLi) alloy melts and reacts in the liquid state with the B powders; after one hour the temperature was raised and maintained at 1000°C for 100 hours. This procedure gave compact cylindrical shaped samples (12 mm diameter and about 10 mm height). The analysis of X-ray diffraction measurements carried out on the top and on the bottom of the cylinder emphasized a little gradient of doping along the height, being the top of the sample richer of Mg than the bottom [29].

Elemental analyses by atomic absorption spectroscopy were performed on two samples with nominal composition $Mg_{0.9}(AlLi)_{0.1}B_2$ and $Mg_{0.7}(AlLi)_{0.3}B_2$ to verify if, during the process, any material loss occurred. The analyses showed that, within the experimental uncertainty, the atomic percentage contents of Mg, Al, Li were equal to the nominal ones.

This preparation technique, developed for pure $MgB_2$, was successfully used to produce $Mg_{1-x}Al_xB_2$ samples with negligible amount of phase separations (less than 3% in volume in the x=0.3 sample) [30] and excellent transport and superconducting properties [29]. The present Al-Li doped samples show a morphology quite similar to those Al-doped samples with a network of well connected grains (1-3 µm size) and no evidences of phase separation. The scanning electron microscopy (SEM) image of a typical Al-Li doped sample is shown in Fig. 1.

**3. Crystallographic characterization**

The characterization of the structure of the present samples was carried out by neutron and XRD diffraction which are discussed in the sections 3.1 and 3.2 respectively.

The details of the procedure adopted to estimate the site occupancy of the Li atoms is reported in the Appendix.

**3.1 Neutron diffraction**

The identification of the site occupancy by Li atoms, which is rather difficult by X-ray diffraction because of the low atomic scattering power of Li, was carried out by neutron diffraction. This technique turned out to be particularly powerful because of the contrast offered by the negative scattering length of Li [$b_{Li}$ = -1.90(3) fm] against the positive values for Al [$b_{Al}$ =3.449(5) fm], Mg [$b_{Mg}$ = 5.375(4) fm] and $^{11}$B [$b_B$ = 6.65(4) fm]. Moreover the use of the non-absorbing $^{11}$B isotope made most favourable the possibility of observing the Li atoms against the other components of the alloys. Neutron diffraction analysis were performed on a sample series $Mg_{1-x}(AlLi)_x{}^{11}B_2$ with x=0. 0.1,0.2,0.3. The experiment was carried out at the high intensity powder diffractometer D20 installed at the HFIR of the ILL (Grenoble, France). A wavelength λ = 1.322 Å was selected for the monochromatic incoming beam and the diffraction pattern was collected over the scattering angle range 10° < 2ϑ < 100° with 0.1° step. The samples were contained inside a vanadium cell with 0.2 mm wall thickness and a standard cryostat was employed for the low temperature runs. The attenuation was separately measured by transmission measurements on the same samples used for the diffraction experiment. As an example, the diffraction pattern measured at 20 K on the x = 0.2 sample is shown in Fig.2 in comparison with the spectrum collected on a pure $MgB_2$ sample under the same experimental conditions. As clearly apparent from the low angle region of the data, a second crystallographic phase, other than the C32 phase which is characteristic of $MgB_2$, develops in the alloy. This second phase, which was observed in all the samples but the $MgB_2$, is probably due to some Li-B compounds but it was impossible to identify it unambiguously because of the lack of literature data (see section 3.2). The second phase amount - evaluated by the integrated intensities of the second phase in respect with those of the C32 phase - comes out roughly to be of the order 5% in volume in the most substituted samples. Luckily enough, this second phase represents a minor contribution to the structure and we could focus on the dominant phase to define the Li concentration and occupancy. The diffraction data were analysed by selecting some specific Bragg reflections whose integrated intensities were carefully studied. The use of integrated intensities ensured a higher statistical accuracy on the results, thus counterbalancing the rather large neutron absorption correction due to natural Li. Fig 3 shows the (100) and (002) Bragg peaks measured on the samples with x=0, 0.1, 0.2, 0.3. An accurate analysis of the integrated intensity of these peaks was carried out as a function of the Al-Li content of the compounds. This set of planes was selected because of their minimal contamination with the minority phase and, for the (100) reflection, the highest sensitivity to the Mg site occupancy. A comparison between the (101) and (002) reflections, which are not strongly affected by the Mg site occupancy, indicated that no preferential orientation was present. After correcting the integrated intensity data for absorption, the crystallographic quantity was obtained from the intensity ratios and, in turn, the scattering length at the Mg site was deduced. The analysis revealed that the B site is fully occupied by boron, while the scattering length at the Mg site, $b_{Mg}^s$, was found to depend on the alloy composition. In particular the $b_{Mg}^s$ value estimated in the undoped sample coincides within the error with the expected value indicating that Mg site is fully occupied. The behaviour of the Mg site scattering length ($b_{Mg}^s$) versus x is shown in the Fig 4: it decreases rather linearly as far as x increases.

If Li and Al enter in the Mg site in the nominal amount, $b_{Mg}^s$ should vary as $b_{Mg}^s = (1-x) \cdot b_{Mg} + \frac{x}{2} \cdot [b_{Al} + b_{Li}]$: this corresponds to the dash-dotted line in Fig 4 which is incompatible with the experimental results. On the opposite, if Li is not entered at all, $b_{Mg}^s$ should vary as $b_{Mg}^s = (1-x) \cdot b_{Mg} + x \cdot b_{Al}$: this corresponds to the dashed line in Fig 4 which clearly does not match the experimental data. The experimental data show that Li populate the Mg site even if less than the nominal amount. The evaluation reported in the Appendix shows that the composition dependence of the Mg site scattering length is compatible with an Al to Li ratio roughly equal to 2:1 .

**3.2 X-ray diffraction**

X-ray powder patterns were collected on all the samples by a Guinier-Stöe camera using the Cu Kα radiation and pure Si as an internal standard (*a*= 5.4308 Å).

XRD patterns for a series of Al-Li doped samples (x=0.1, 0.2, 0.3, 0.35 and 0.45 with natural boron) are shown in Fig. 5(panel (a)). Peaks of the doped compounds are shifted in comparison with pure MgB$_2$ peaks and they broaden on increasing the dopant content showing that the hexagonal phase becomes less and less stable (see in particular the shift and broadening of (110) and (002) reflections showed in Fig. 5(b)).

Fig.6 shows the lattice constants of the Mg$_{(1-x)}$(AlLi)$_x$B$_2$. Like in Al doped compounds, *a* and *c* axes decrease as far as the substituted Mg increases: *a* axis shows a little variation (less than 0.6%) while *c* axis varies as much as ~2-3 %.

As emphasized by neutron diffraction, in the low angle region of the alloy data, XRD peaks of some minority spurious phase are present. Neutron diffraction analysis suggests that they could belong to a Li-B phase. To verify this hypothesis a sample of nominal composition "LiB$_2$" was purposely prepared with the same procedure as described in section 2. The XRD pattern of this compound is shown in Fig. 7: at low angles (10° < 2θ < 35°) it presents strong reflections which do not correspond to any other of the already known Li-B compound. In the same figure, a magnification of the XRD pattern of the Mg$_{0.7}$(AlLi)$_{0.3}$B$_2$ compound is shown. It is evident that all the peaks of the "LiB$_2$" compound correspond to the spurious peaks of the Mg$_{0.7}$(AlLi)$_{0.3}$B$_2$ compound. So we identify the minority phase present in our samples as a binary Li-B compound. Considering the relative integrated intensities of MgB$_2$ and "LiB$_2$", we can give a rough estimate of this second phase which should not exceed 5-7 % in volume in agreement with the evaluation given by neutron diffraction.

**4. Electrical and magnetic characterization**

DC susceptibility measurements were performed with a SQUID magnetometer (MPMS Quantum Design) under a magnetic field of 10 Gauss. Susceptibility measurements from 5 to 45 K are shown in Fig.8a for samples of different nominal composition, while Tab.I summarizes results about a series of samples. The critical temperature $T_c^\chi$ is evaluated as the mean value between the temperature respectively at the 90% and 10% of the transition; $T_c^\chi$ decreases monotonically from 38.5K for the pure sample to 18.1K and the transition width $\Delta T_c^\chi$ ( defined as the temperature difference between the 90% and 10% of the transition) becomes larger for the most heavily doped samples. The low temperature susceptibility values in Fig. 8 are normalized to -1 (complete flux espulsion). For all the samples with geometry regular enough to evaluate precisely the volume, the low temperature values were, within the experimental errors, close to -1, confirming the absence of remarkable amount of not superconducting spurious phases.

To verify the reproducibility of our results, several samples with the same nominal composition were prepared. In Fig. 8b the susceptibility of four different samples with the same nominal composition $Mg_{0.8}(AlLi)_{0.2}B_2$, are shown (two samples were prepared with natural boron, the other two with isotopically enriched $^{11}B$). The transition is rather sharp in all the samples, and the transition temperatures of the four samples vary within 2 K. Similar analysis on samples with $Mg_{0.9}(AlLi)_{0.1}B_2$ nominal composition gives a $T_c$ spread of about 1 K. This corresponds to a relative error on the substituted Mg of about 10% that can be accounted for by uncertainty in the nominal composition as well as the gradient of doping that we observe in bulk samples as discussed in section 2.

Resistivity was measured with a four probe technique by a PPMS QD. Resistivity measurements as a function of temperature from 20 to 300 K are shown in Fig.9a for Al-Li doped compounds. In Fig.9b the temperature range around the transition temperature is enlarged. The transition temperature decreases monotonically with increasing the doping and no multiple transitions are present. Residual resistivity $\rho(40)$ increases with doping from 2.5$\mu\Omega$cm up to 40$\mu\Omega$cm (for x=0.45). The low $\rho(40)$ values indicate that grain boundaries resistivity is negligible in our samples and their monotonous increase with doping indicate that Al-Li substitutions are effective in reducing the electron mean free path. The residual resistivity ratio (RRR) varies from 7.2 for the pure sample to 1.5 for heavily doped samples. Similar values were obtained in the Al-series [29]. The critical temperatures $T_c^\rho$, estimated as the mean value between the temperature respectively at the 90% and 10% of the resistive transition, are systematically higher than the ones measured by susceptibility reaching the value of 25.6K for x=0.45. The transition width ΔT$_c$ is estimated as the difference between the temperature respectively at the 90% and 10% of the resistive transition and becomes as large as 5 K at the highest doping.

**5. Specific heat**

Specific heat measurements were performed from 2 to 300 K on three Al-Li doped samples (x=0.1, 0.2, 0.3) by a PPMS-7T QD using the relaxation method. For each Al-Li concentration we performed a set of measurements between 2 and 40 K and in magnetic field of 0, 3, 5 and 7 T. Results are plotted as c/T-vs-T$^2$ in Fig. 10 where, for sake of clarity we plot only data in zero field and 7 T. The superconducting contributions can be easily visualized by comparing data in zero field with those obtained in magnetic field: the jump at T$_c$ which is well pronounced in undoped MgB$_2$ (see ref. [9]), gets broader as the Al-Li concentration increases. T$_c$ was defined at half the specific heat anomaly and the uncertainty takes into account the transition broadening; the dependence of T$_c$ on the Al-Li concentration x is reported in Table II. T$_c$ values monotonously decrease as the doping increases as observed in the previous section by resistivity and susceptibility measurements. It is worth noting that, as usual, the T$_c$ values estimated from specific heat measurements are lower than those estimated from resistivity and susceptibility (see table I).

From data obtained at 7 T, the temperature dependence of the normal state specific heat can be evaluated. A close data inspection reveals that Schottky anomalies are very small for x=0.1 and become observable below 5 K for x=0.2 and x=0.3. Neglecting these low temperature data, the curve obtained with B = 7 T were fitted from T$_c$(7T) up to 40 K by the model function $c(B=7T) = \gamma \cdot T + \beta \cdot T^3 + \delta \cdot T^5$ where γ is the Sommerfeld coefficient and the T$^5$ term was introduced as a high temperature correction to the T$^3$ Debye law. Results are reported in Table II. As check of the reliability of the normal state parameters, we estimated the superconducting electronic contribution $c_{sc}$ as $c_{sc} = c(B=0T) - [\beta \cdot T^3 + \delta \cdot T^5]$. The entropy difference $\Delta S(T) = \int_0^T (c_{sc}/T - \gamma) \cdot dT$ between the superconducting and the normal state contribution was evaluated (see the insets of Fig. 10) and it can be noted that ΔS actually vanishes close to $T_c$ for all the measured samples as required by the entropy balance between the normal and the superconducting state.

The fitting coefficients β, γ, and δ of the normal state curve are reported in Table II for each Al-Li concentration. We found that γ progressively decreases from 3.0±0.2mJ/molK$^2$ for undoped MgB$_2$ to 2.4±0.3mJ/molK$^2$ for Mg$_{0.7}$(AlLi)$_{0.3}$B$_2$. A reduction of γ as a function of Mg substituted was observed also in Al-doped samples [9] and this can be related to both a decrease of density of state at the Fermi level and a suppression of electron-phonon coupling. The Debye temperature $\Theta_D$ in the limit of low temperature is calculated by the β coefficients as $\Theta_D = (1944/\beta)^{1/3}$. The x-dependence of $\Theta_D$ is shown in Table II. $\Theta_D$ progressively decreases as the doping increases from 670 K to 580 K. This is probably due to the softening of the phonon modes induced by the doping.

Finally, it is worth noting that specific heat provides a further check on the sample quality. Even small amounts of spurious phases are indeed expected to give a visible contribution to the heat capacity, since MgB$_2$ has low specific heat at low temperature (MgB$_2$ is a superconductor with high Debye temperature). The data analysis evidenced that the entropy balance between the superconducting and the normal electronic state vanishes close to $T_c$ and only little Schottky anomaly are visible in data of sample with x=0.2 and 0.3 consistently with the fact that spurious phases are present only in small amount. Moreover, the specific heat values of the most doped sample at 40 K are only 10% higher than those of the undoped one (compare fig.10 with fig.2 of ref. 9) and this difference can be reasonably taken into account by a reduced $\Theta_D$.

**6. Raman Spectroscopy**

Raman spectra were collected using a microRaman spectrometer Infinity by Jobin Yvon. The spectrometer, equipped with a He-Ne laser source, works in back-scattering geometry and uses a CCD (charge coupling device) detector to collect the scattered light dispersed by an 1800 line/mm grating. A notch filter was used to reject the elastically scattered light. The microscope, equipped with a 20X objective, allowed us to achieve a spatial resolution of a few microns on the sample surface. For each sample Raman spectra were collected over the expected phonon frequency range (i.e. 200-1100 $cm^{-1}$) from several sample regions over freshly cut surfaces. The spectra collected from from the different sampled areas show similar spectral features, thus suggesting a good homogeneity degree of the samples at least over a micrometric scale. All the spectra were analyzed by a standard fitting procedure [5, 31]: a fitting function given by a linear combination of damped harmonic oscillator plus an electronic term has been used to account for the phonon structures and the diffusive electron scattering term respectively [31]. For each sample, both the direct comparison among the spectra and the rather close best fit values for the phonon peaks obtained from different regions made us confident about the good sample homogeneity. The best-fit values of phonon frequencies and widths were then averaged and their standard deviations were used as an estimate of the experimental uncertainty.

Raman spectra of Al doped samples are shown in panel (a) of Fig. 11. Our data are in a good agreement with previous experimental results [5, 7] and confirm that the Al-doping strongly affects the Raman spectrum which, in the pure compound, is characterized by a very broad peak centred at around 600 $cm^{-1}$, usually ascribed to the $E_{2g}$ phonon mode. The Raman spectrum of Al-doped compounds shows a high frequency two-peak structure, absent in the undoped compounds, whose frequencies and intensities increase with the Al content [5, 7]. Although a definite assignment of the spectral features of the Raman spectrum has not been agreed upon yet [32-34], it has been suggested that the doping dependence of high-frequency structures should be related to a progressive reduction of the electron-phonon coupling and the consequent mode frequency renormalization [5].

A similar behaviour is shown by the Raman spectra of the Al-Li doped samples shown in panel (b) of Fig. 11. It is worth to notice that the factor group analysis predicts four modes at the Γ point of the Brillouin zone for the hexagonal $MgB_2$: $E_u$, $A_{2u}$, $E_{2g}$ $B_{1g}$, where only the $E_{2g}$ mode is Raman active. The clear evidence of more than a single phonon peak in doped samples is thus an apparent violation of the selection rules. This is not unusual when dealing with disordered and/or defective systems where a relaxation of the momentum-selection rule may occur, leading to Brillouin zone folding and consequently to the appearance in the Raman spectrum of additional features connected with phonons lying beyond the zone centre. In these systems the Raman spectrum can thus reflect the features of the whole phonon density of state. This seems to be the case of $MgB_2$-doped compounds where the introduction of on-site disorder activates Raman-forbidden modes and makes the spectrum more similar to the phonon density of states [5] . This idea is supported by recent Raman data on C-doped $MgB_2$ where spectra similar to the present ones have been obtained for very small concentration of substituted B [34] Under this hypothesis the present results suggest that the whole lattice dynamic appears to be affected by the Al doping only.

**7. Discussion and conclusion.**

The present analysis of the diffraction data has shown that the codoped compounds have the stoichiometry $Mg_{(1-x)}(Al_\alpha Li_{1-\alpha})_x B_2$, with $\alpha \sim 2/3$. Let us assume that the valence of each dopant can be summed to give the total charge introduced per cell. So codoping with Al and Li should leave an unbalanced number of electrons per cell equal to $[\alpha - (1-\alpha)] \cdot x \approx \frac{1}{3} x$, that is the same amount $x$ of substituted Mg induces an effective charge doping $x$ for Al-doped and only $x/3$ for our Al-Li codoped samples.

Fig. 12 shows the lattice parameters of $Mg_{(1-x)}(Al_\alpha Li_{1-\alpha})_x B_2$ in comparison with those of $Mg_{(1-x)}Al_x B_2$ prepared with the same technique [29]. For both series, the $a$ and $c$ axes decrease as a function of $x$ (Fig.12a) although a smaller lattice compression is observed for the Al-Li series. In Fig. 12b the lattice constants of the two series are plotted as a function of the Al content, which in the Al-Li codoped samples is simply given by $\frac{2}{3}x \pm \Delta\alpha \cdot x$. In this case, an almost complete overlap of the lattice parameters of the two series is obtained. The success of the Al-scaling suggests that the Al content, regardless of Li, causes the volume compression. This result is consistent with the Li atomic radius being very similar to that of Mg.

The $T_c$ values of Al-Li and Al doped samples, plotted as a function of different variables, are compared in Fig.13. We point out that the critical temperatures of the two series are evaluated resistively, under the same experimental condition and $T_c$ definition. With increasing $x$, $T_c$ decreases almost linearly in both the sample series. Fig. 13a shows $T_c$ versus the effective charge doping ($\frac{1}{3}x \pm 2\Delta\alpha \cdot x$ and $x \pm 0.1x$ for the Al-Li and Al respectively). For given charge doping value, the Al-Li samples exhibit $T_c$ values significantly lower than those of the Al doped ones, suggesting that effects additional to charge doping induce a further suppression of the superconductive phase. In Fig. 13b, $T_c$ is plotted as a function of $x$: in this case, the Al doped samples present a lower $T_c$. Finally, in Fig. 13c where $T_c$ versus the Al content is shown, the data from the two series scale on the same curve. This findings supports the idea that the superconductivity is much more affected by the lattice distortion induced by Al substitution than by the band filling, and, as a first approximation, the presence of Li does not affect the critical temperature.

To enlighten the connection between lattice deformations and superconductivity, $T_c$ versus the c-axis is shown in Fig.14 where the data from Al-Li and Al series fall on the same curve at least for high c-axis values (i.e. low doping). The Al doped series show a $T_c$ slightly but systematically lower than the Al-Li series at low c-axis values (i.e. high doping). Bearing in mind that the samples of the Al series are doped with an effective charge larger than the Al-Li series, the above discrepancy may be ascribed to charge effect which becomes more important at high Al concentrations (>30%), where $T_c$ vs Al concentration decreases faster [18,19,21].

In Fig. 15 we compare the Sommerfeld coefficient γ of the $Mg_{(1-x)}(AlLi)_xB_2$ with those of $Mg_{(1-x)}Al_xB_2$ (from ref. [9]). As done before, we plot γ –versus- charge doping (Fig. 15a), -$x$ (Fig. 15b) and -Al content (Fig. 15c). Also in this case, notwithstanding the large experimental uncertainty, the better scaling parameter is the Al content: γ linearly decreases from 3 to 2.1 mJ/mol K$^2$ with Al content increasing from 0 to 0.3.

The γ value is proportional to $N(0)(1+\bar{\lambda})$ where $N(0)$ is the density of states at the Fermi level, and $\bar{\lambda}$ is the electron-phonon coupling constant averaged on the σ and π bands. For the charge doping level we considered (lower than 0.3 electron/cell) the maximum diminution of $N(0)$ is 16% [19]; on the contrary, γ decreases by up to 30%, which implies a remarkable the reduction of $\bar{\lambda}$. In fact, assuming $\bar{\lambda}$ =0.8 for undoped $MgB_2$ it decreases down to 0.55 increasing the Al content up to 0.3. We can conclude that Al content in both the sample series causes a remarkable suppression of the electron phonon coupling.

Finally we compare the Raman phonon spectra of the two sample series. In Fig. 16a two spectra at the same Al content (20%) but different percentage of substituted Mg (30% for the Al-Li sample and 20% for the Al sample) are shown. The remarkable similarity between the spectral structures of the two spectra suggests that also the phonon spectrum, and thus the lattice dynamics, is mainly affected by the presence of Al. This idea is confirmed by looking at Figs. 16b and 16c, where the frequencies of the main structures of the spectrum versus charge doping and Al content are shown for both Al and Al-Li doped samples. Also in this case, the better agreement is obtained when the Al dependence is chosen, whereas the presence of Li seems to be actually ineffective. Bearing in mind that the electron-phonon coupling is the underlying mechanism for the superconducting behaviour, Raman results support the hypothesis of a major role of Al- against Li-doping in tuning the superconducting properties of $MgB_2$.

The overall experimental results we presented depict a consistent frame. In both Al and Al-Li doped compounds the substitution of Al at the Mg-sites drives the suppression of superconductivity. The reduction of $T_c$ is accompanied by a reduction of electron phonon coupling as suggested by the Raman results and estimated by the analysis of the Sommerfeld coefficient.

The novel result which emerges consistently from the comparison between the two series of compounds is the estimate of the relative importance of the main effects induced by the presence of Al in the $MgB_2$ lattice, namely point-like defects, lattice compression and σ-band filling. Since charge doping should be substantially suppressed in our Al-Li compounds, and Al content was proved to be the leading parameter for the scaling of all the properties measured in Al and Al-Li series, our results point out at a major role of the lattice deformations against charge doping. More precisely, our data suggest that charge doping mechanism *alone* is not the most effective in tuning the superconducting properties. On the other hand, lattice compression raises the Fermi level filling the σ bands and substitutional defects produces intra-band and inter-band scattering. The role of the latter mechanism has been emphasized [20] while the importance of the former is well proved by the suppression of superconductivity in irradiated $MgB_2$ [24-26], where neither charge doping nor lattice compression occur. These effects, *all together*, can influence in substantial way the superconductivity in Al doped compounds.

As stated before these conclusions are subjected to the hypothesis that the valence of each dopant can be summed, that is to say lithium is able to compensate the charge introduced by aluminum. If this is not the case the fact that the charge is not a good scaling parameter will be not surprising and our results should make everyone careful about summing the charge introduced by elements, even in comparing substitutions in the same crystal site. Even more care should be used in comparing substitutions in different site, such as that of Al in the Mg site with that of C in the B site.

In conclusion, the present paper reports on a detailed and extended investigation of a series of Al-Li codoped $MgB_2$ samples prepared following a procedure already successfully applied to produce high-quality pure and Al-doped $MgB_2$ samples. The quality of the presently investigated samples was controlled by means of an exceptionally extended characterization procedures which exploited a large number of experimental techniques and was mostly aimed at verifying the effective amount of substitutional Li in the $MgB_2$ lattice. The results here obtained show an effective Al-Li ratio close to 2:1 and a minority phase of a binary Li-B compound (about 5% in the most doped sample), as detected by diffraction techniques. On the other hand no evidences of significant secondary phases have been obtained by the specific heat measurements. Finally susceptibility and resistivity measurements have shown a nearly complete flux expulsion, the absence of multiple transitions, and a monotone decrease of $T_C$ with the doping.

The whole of the results, when compared with those obtained on the homologous series of $Mg_{1-x}Al_xB_2$ samples, shows that Li has almost no effect on the superconducting properties. In our opinion for a more realistic comprehension of the phenomena, at least in the presently exploited doping range, a detailed analysis of all the effects introduced by Al and Li should be envisaged.

**Acknowledgements**

Dr. Thomas Hansen from Institut Laue-Langevin (Grenoble, France) is kindly acknowledged for providing beam time at the high-flux diffractometer D20. Some of the authors are partially supported by Ministry of Italian Research by a PRIN2004 project.

**Appendix: Li occupancy estimate.**

The analysis of the scattering length at the Mg site emphasized that Li populate the Mg site even though at a lower extent than the nominal amount (see section 3.1). This implies the formation of a secondary phase that was identified as a Li-B phase whose content was estimated to be less than 10% in volume. Starting from these data the amount of Li in the substituted site can be determined as follows.

We consider the generic C32 phase $Mg_{1-x_A}(Al_\alpha Li_{1-\alpha})_{x_A} B_2$ where the Li occupancy is not fixed (being equal to $1-\alpha$) and we introduce $x_A$ as the actual substituted Mg. Due to the presence of the spurious phase $x_A$ can be different from the nominal $x$. So $b^s_{Mg}$ varies as:

$$b_{Mg}^{s} = (1-x_A) \cdot b_{Mg} + x_A \cdot [\alpha \cdot b_{Al} + (1-\alpha) \cdot b_{Li}] \tag{1}$$

Since the elemental analysis shows that there are no material losses during the sample preparation (see section 2), we can write down the balance equation:

$$Mg_{1-x}(Al_{0.5}Li_{0.5})_x B_2 = (1-y)[Mg_{1-x_A}(Al_\alpha Li_{1-\alpha})_{x_A} B_2] + y(Li_b Mg_c Al_d B_2)$$

(2)

The left-hand side corresponds to one mole of C32 phase with the nominal composition. The right-hand side represents the result of the reaction: a molar fraction *1-y* of a C32 phase and a molar fraction *y* of spurious phases. *b, c* and *d* are, respectively the amount of Li , Mg,  and Al  in the spurious phases (*b+c+d=1*). A value of α different from 0.5 implies the presence of some spurious phases to keep constant the total amount of each element.

Now we want to estimate α: this is accomplished performing a best fit procedure of the experimental data by means of eq. (1) with α as the fitting parameter, provided that the constraints imposed by eq (2) are satisfied.

Starting from the simplest hypothesis $x_A=x$, we obtain α=0.72. This value is quite unlikely because an amount of spurious phase much larger than our previous evaluation must be hypothesized (*y*=0.30 for the *x* =0.3 sample), to compensate such an excess of Al content in the C32 phase. However, it is not necessary to set the constrain, $x_A=x$, indeed starting from eq. (2) and writing down three balance equations for Li, Mg and Al, we can solve them and perform the best fit procedure for each given composition of spurious phase (i.e. each "c,d" assigned value in eq (2)) obtaining α, $x_A$ and *y* as a function of "c,d".

Following the indication of the X-ray analysis which identifies a Li-B phase we impose the constraint c=d=0   and obtain α=0.60 (with y=0.05 in the x=0.3 sample).  Even if the spurious phase contains a minor amount of Mg and Al (i.e. setting $c+d \leq 0.2$ ) α vary from 0.58 to 0.61 and *y*

from 0.06 to 0.07.

Even considering every possible "c,d" values, α never exceeds 0.72, while if the total amount of spurious phases is restricted imposing $y \leq 0.15$ in the x=0.3 sample, α never exceeds 0.62.

In the end, we are confident that a value of α ranging from 0.6 to 0.72 is highly possible, being the lower more reliable than the upper limit of the interval.

As already noted, $x_A$ can be different from $x$, in particular it is slightly smaller than $x$. It comes out that $x_A=x$ is of the same order of magnitude as arising from the preparation technique (see also [25]) and does not affect the discussion (for example in the $x$=0.3 sample $x_A=x \leq 0.05$). As a result we have taken $x_A=x$.

From the whole of these results we can conclude that the composition dependence of the Mg site scattering length is compatible with an Al to Li ratio roughly equal to 2:1 (α~2/3).

**Figure captions**

**Figure 1**: SEM image of $Mg_{0.8}(AlLi)_{0.2}B_2$.

**Figure 2**: Raw neutron diffraction data versus scattering angle in $MgB_2$ (lower curve) and $Mg_{0.8}(AlLi)_{0.2}B_2$ (upper curve).

**Figure 3**: Intensity of the Bragg peaks (100) (left panel) and (002) (right panel) measured on the samples with x=0, 0.1, 0.2, 0.3.

**Figure 4:** The measured scattering length in the Mg site, $b_{Mg}^s$, as a function of $x$ for various values of $\alpha$.; dash-dotted line corresponds with an equal amount of Al and Li substituted ($\alpha$=0.5); dashed line corresponds with the case that Li is not entered at all ($\alpha$=1).

**Figure 5**: XRD patterns for a series of Al-Li doped samples (panel (a)). The reflections (002) and (110) as a function of doping (panel (b)).

**Figure 6**: Lattice parameter *a* (triangles) and *c* (circles) relative to those at zero concentration ($a_0$=3.085Å; $c_0$=3.525Å) in $Mg_{1-x}(AlLi)_xB_2$.

**Figure 7**: XRD patterns of a sample of nominal composition "$LiB_2$" (upper curve) and of the $Mg_{0.7}(AlLi)_{0.3}B_2$ compound (lower curve).

**Figure 8**: Normalized DC susceptibility $\chi$ vs temperature for a series of $Mg_{1-x}(AlLi)_xB_2$ with x=0, 0.05, 0.1, 0.2, 0.3, 0.35, 0.4, 0.45 (panel (a)) and with x=0.2 (panel (b)).

**Figure 9**: Resistivity as a function of temperature for the $Mg_{1-x}(AlLi)_xB_2$ samples with x=0, 0.05, 0.1, 0.2 , 0.3, 0.35, 0.4, 0.45. In the right panel the resistive transition are shown in detail.

**Figure 10**: Low temperature specific heat of $Mg_{1-x}(AlLi)_xB_2$ with $x$= 0.1, 0.2, 0.3 (from the top to the bottom) plotted as c/T vs-$T^2$. For sake of clarity, only measurements in zero field (filled squares) and in 7 T (open squares) are reported for each sample. Insets: The entropy difference $\Delta S$ calculated by integrating $c_{sc}/T - \gamma$ with respect to temperature.

**Figure 11**: Raman spectra of the Al (panel a) and Al-Li (panel b) doped samples for different value x of the substituted Mg. The full lines are fit to the experimental data. In the right panel the Raman spectrum of pure $MgB_2$ is shown for comparison.

**Figure 12**: Lattice parameter *a* (triangles) and *c* (circles) relative to those at zero concentration ($a_0$=3.085Å; $c_0$=3.525Å) in $Mg_{1-x}(AlLi)_xB_2$ (empty symbols) and $Mg_{1-x}Al_xB_2$ (filled symbols) [25] as a function of x (panel (a)) and Aluminium content (panel (b)).

**Figure 13:** $T_c$ in $Mg_{1-x}(AlLi)_xB_2$ (empty symbols) and $Mg_{1-x}Al_xB_2$ (filled symbols) as a function of charge doping (a), x (b) and Al content (c).

**Figure 14**: Critical temperature in $Mg_{1-x}(AlLi)_xB_2$ (empty symbols) and $Mg_{1-x}Al_xB_2$ (filled symbols) as a function of *c* axis.

**Figure 15**: Sommerfeld coefficient γ in $Mg_{1-x}(AlLi)_xB_2$ (empty symbols) and $Mg_{1-x}Al_xB_2$ (filled symbols) as a function of charge doping (a), x (b) and Al content (c).

**Figure 16**: (a) Raman spectra of $Mg_{0.7}(AlLi)_{0.3}B_2$ and $Mg_{0.8}Al_{0.2}B_2$. The frequencies $\nu_1$, $\nu_2$ and $\nu_3$ indicated in the panel (a) as a function of ED (b) and Al content (c).

**Table captions**

**Table I**: Parameters of a selected series of $Mg_{1-x}(AlLi)_xB_2$ polycrystalline samples: the crystallographic *a* and *c* axes; the critical temperatures $T_c^\chi$ (and $T_c^\rho$) defined as $T_c=(T_{90\%}+T_{10\%})/2$; the transition widths $\Delta T_c = (T_{90\%}-T_{10\%})$ where $T_{90\%}$ and $T_{10\%}$ are estimated at the 90% and 10% of the susceptibility (resistive) transition; $\rho_0$ defined as the resistivity measured at 40 K; the residual resistivity ratio defined as RRR=*ρ(300)/ρ(40)*.

**Table II:** The critical temperature $T_c$, the γ (Sommerfeld constant), β and δ fitting coefficients of the normal state curve $c(B = 7T) = \gamma \cdot T + \beta \cdot T^3 + \delta \cdot T^5$ and Debye temperature $\Theta_D$ for different Al-Li concentrations *x* in $Mg_{1-x}(AlLi)_xB_2$ obtained by the fitting of specific heat data . The parameters obtained for the undoped $MgB_2$ are from ref. [9].

**Table I**

| x | $a$ axis (Å) | $c$ axis (Å) | $T_c^\chi$ (K) | $\Delta T_c^\chi$ (K) | $T_c^\rho$ (K) | $\Delta T_c^\rho$ (K) | $\rho_0$ (μΩ·cm) | RRR |
|---|---|---|---|---|---|---|---|---|
| 0 | 3.085 | 3.525 | 38.5 | 0.8 | 38.8 | 0.4 | 2.5±0.3 | 7.2 |
| 0.05 | 3.083 | 3.512 | 36.2 | 0.6 | 37.0 | 0.7 | 6.5±0.4 | 3.3 |
| 0.1 | 3.082 | 3.497 | 34.1 | 1.2 | 36.4 | 1.5 | 8.2±0.5 | 2.6 |
| 0.2 | 3.080 | 3.474 | 29.0 | 1.3 | 32.9 | 3.0 | 14±1 | 2.0 |
| 0.3 | 3.074 | 3.454 | 26.9 | 2.2 | 32.0 | 4.0 | 32±4 | 1.8 |
| 0.35 | 3.075 | 3.453 | 25.1 | 2.4 | 30.7 | 5.0 | 38±3 | 1.8 |
| 0.4 | 3.069 | 3.435 | 20.4 | 3.0 | 26.1 | 4.4 | 33±3 | 1.5 |
| 0.45 | 3.070 | 3.432 | 18.1 | 3.6 | 25.6 | 4.1 | 40±3 | 1.5 |

**Table II**

| x | $T_c$ (K) | $\gamma$ (mJ/mol K$^2$) | $\beta$ (mJ/mol K$^4$) | $\delta$ (mJ/mol K$^6$) | $\Theta_D$ (K) |
|---|---|---|---|---|---|
| 0 | 38.0±0.3 | 3.0±0.2 | (6.4±0.2)×10$^{-3}$ | (2.4±0.2)×10$^{-6}$ | 670±15 |
| 0.1 | 31±2 | 2.9±0.2 | (7.4±0.4)×10$^{-3}$ | (1.9±0.2)×10$^{-6}$ | 640±15 |
| 0.2 | 28±3 | 2.7±0.2 | (8.3±0.4)×10$^{-3}$ | (1.9±0.2)×10$^{-6}$ | 615±15 |
| 0.3 | 24±4 | 2.4±0.3 | (10±0.4)×10$^{-3}$ | (1.1±0.2)×10$^{-6}$ | 580±15 |

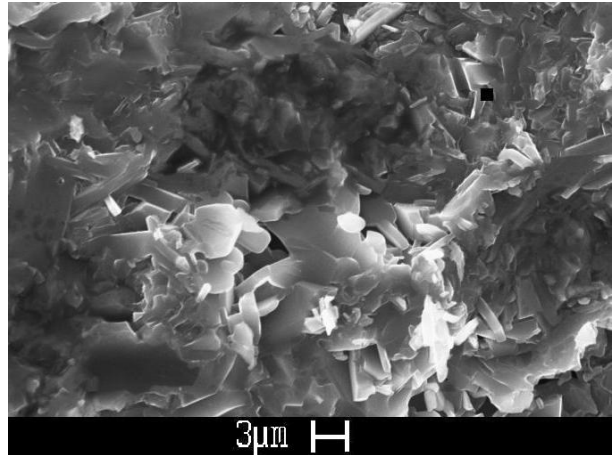

Figure 1

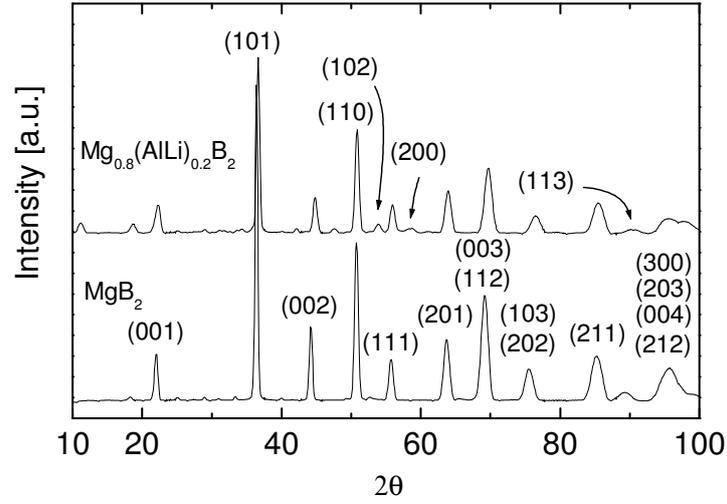

Figure 2

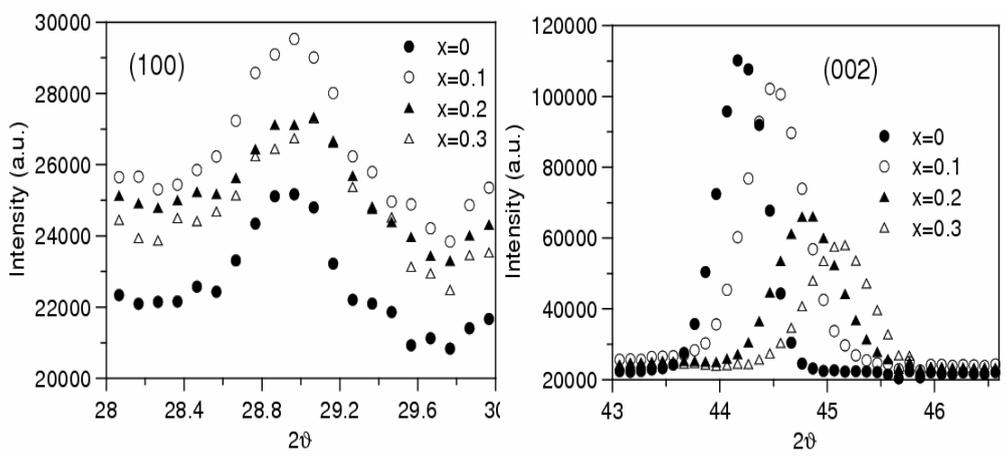

Figure 3

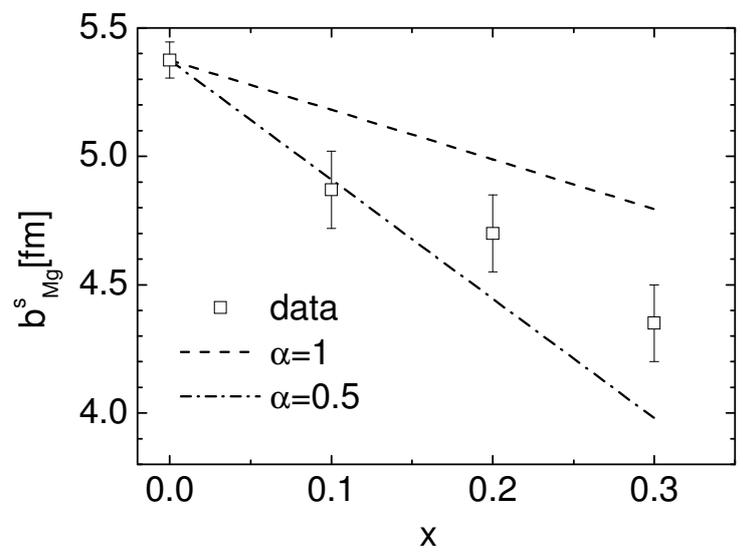

Figure 4

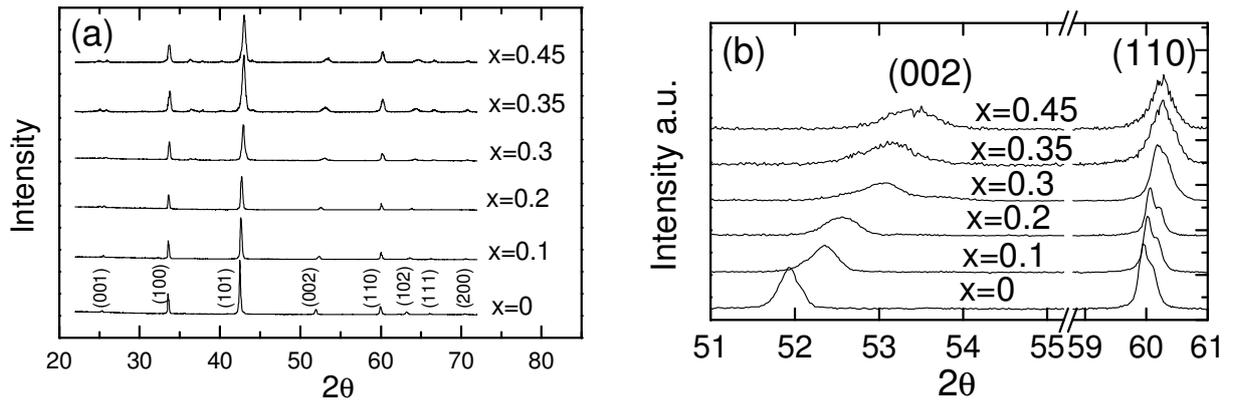

Figure 5

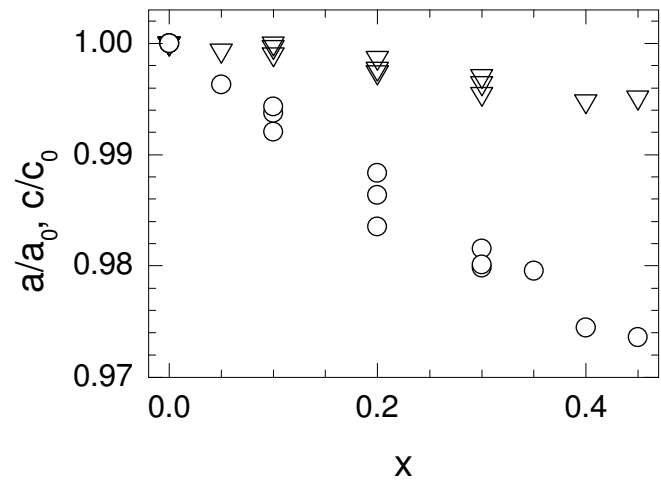

Figure 6

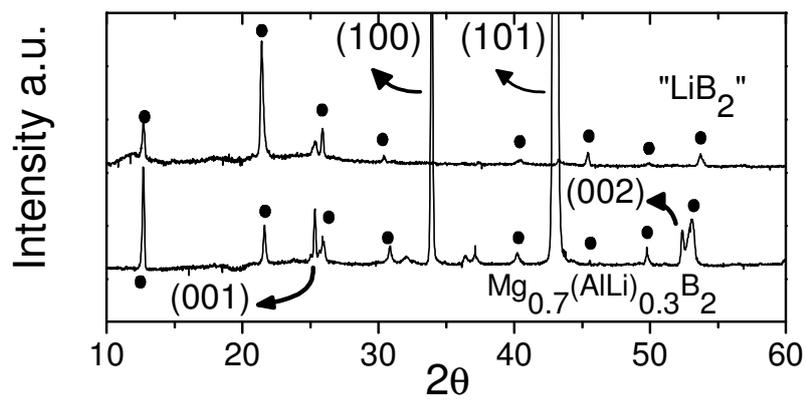

Figure 7

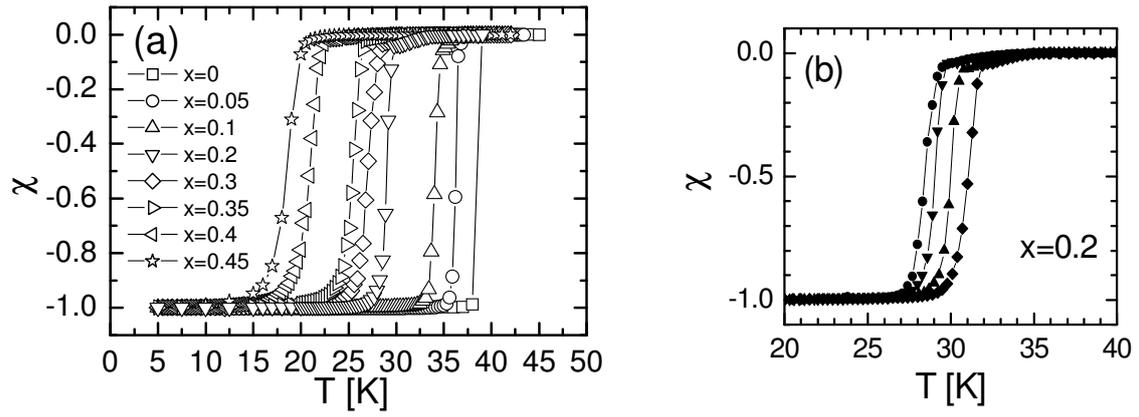

Figure 8

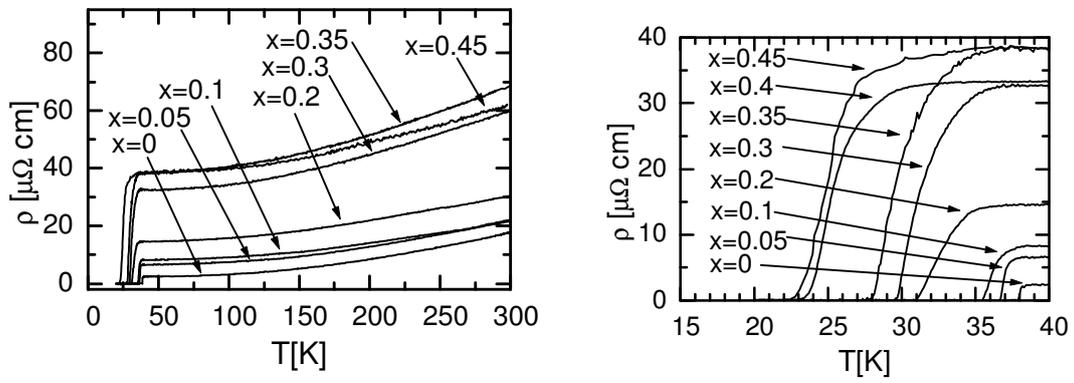

Figure 9

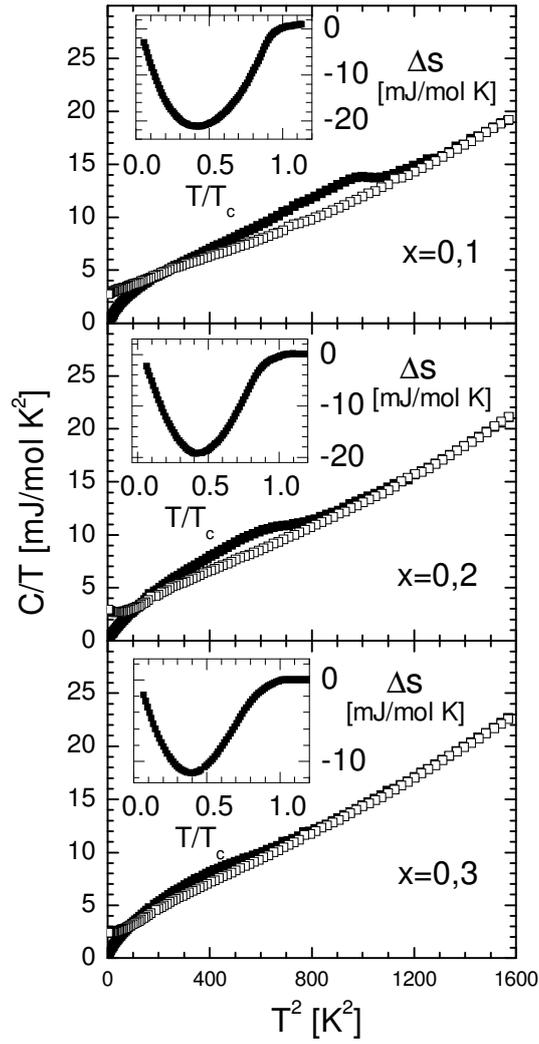

Figure 10

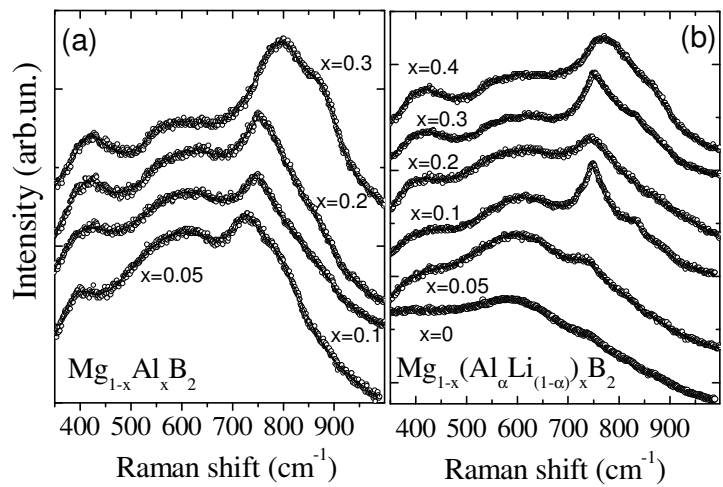

Figure 11.

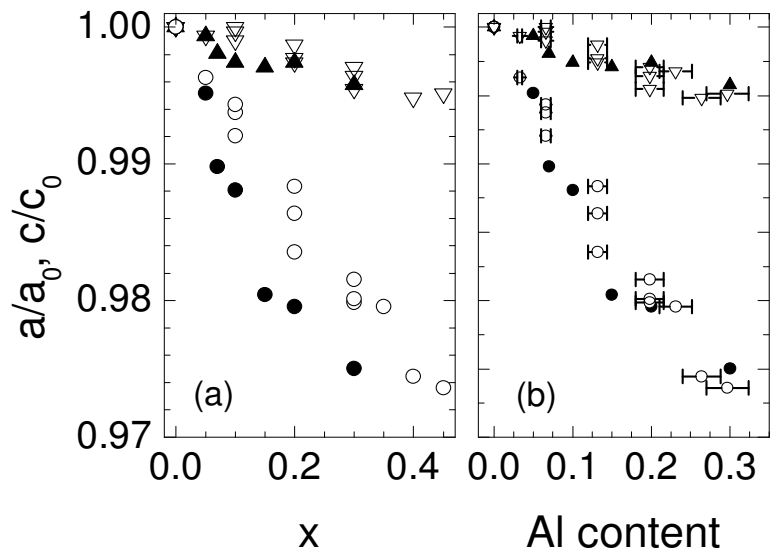

Figure 12

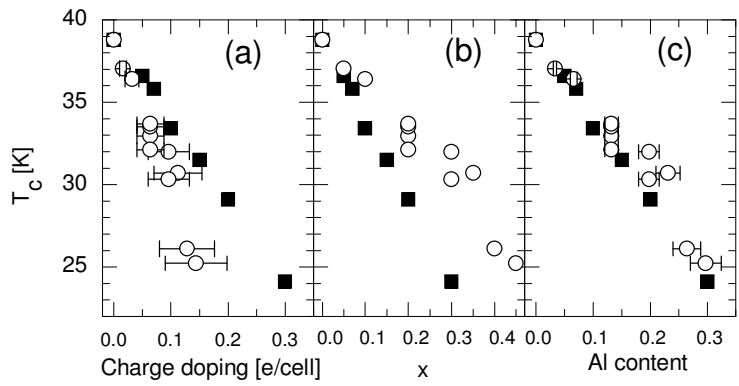

Figure 13

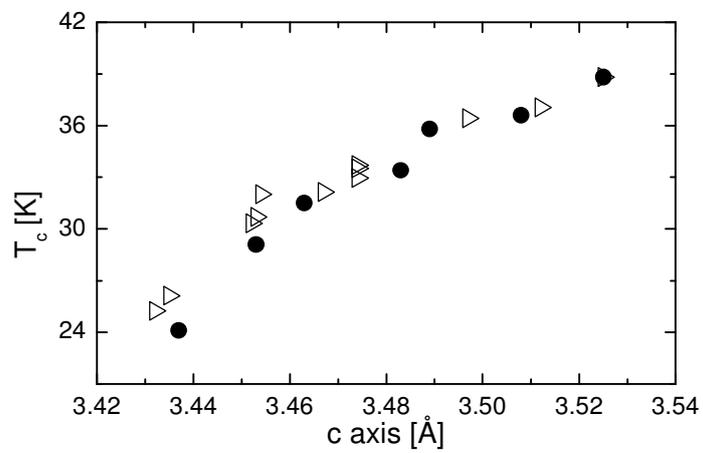

Figure 14

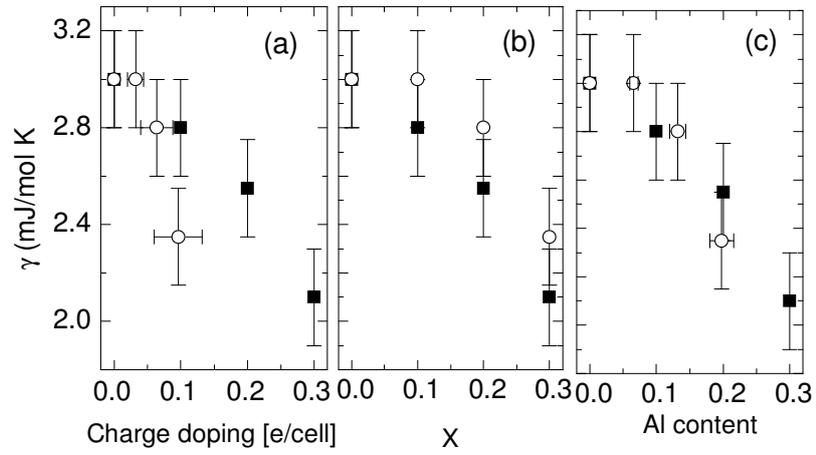

Figure 15

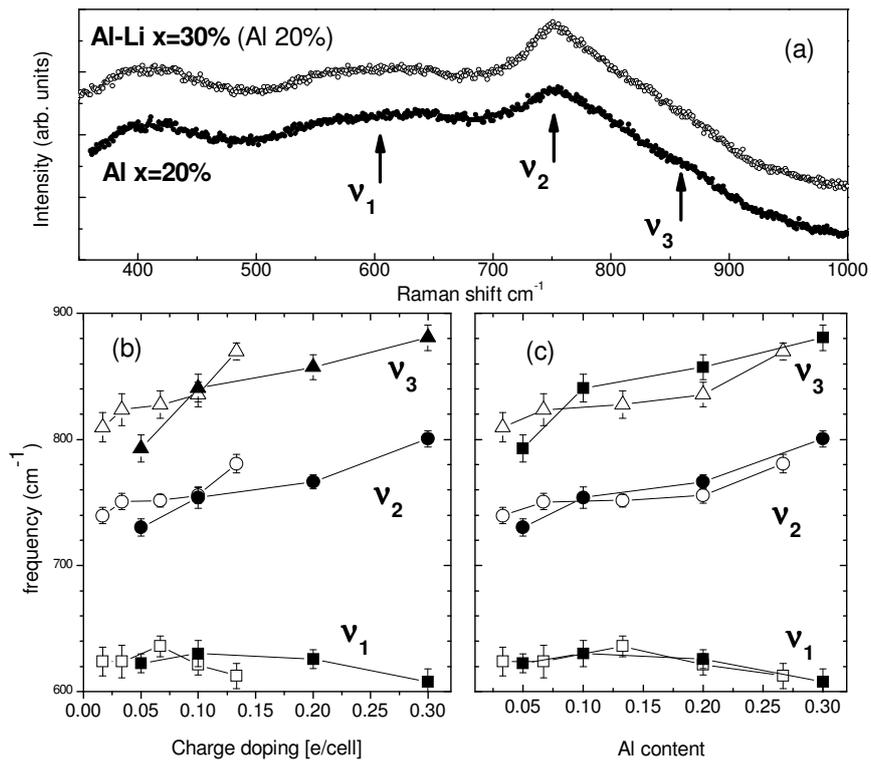

Figure 16